\documentclass[pra,aps,twocolumn,longbibliography]{revtex4-1}

\usepackage{amsmath,amsfonts,amssymb,amsthm,mathtools} 
\usepackage[locale=US]{siunitx}
\usepackage{graphicx}
\usepackage{color}
\usepackage{hyperref}
\usepackage{bm}
\usepackage{float}
\usepackage{placeins}

\begin{document}

\title{Quantum autoencoders to denoise quantum data}

\author{D. Bondarenko}
\affiliation{Institut f\"ur Theoretische Physik, Leibniz Universit\"at Hannover, Appelstr. 2, DE-30167 Hannover, Germany}
\author{P. Feldmann}
\affiliation{Institut f\"ur Theoretische Physik, Leibniz Universit\"at Hannover, Appelstr. 2, DE-30167 Hannover, Germany}

\begin{abstract}
Entangled states are an important resource for quantum computation, communication, metrology, and the simulation of many-body systems. However, noise limits the experimental preparation of such states. Classical data can be efficiently denoised by autoencoders---neural networks trained in unsupervised manner. We develop a novel quantum autoencoder that successfully denoises Greenberger-Horne-Zeilinger states subject to spin-flip errors and random unitary noise. Various emergent quantum technologies could benefit from the proposed unsupervised quantum neural networks.

\end{abstract}

\maketitle
\date{\today}


\section{Introduction}
\label{sec:Intro}
\FloatBarrier
    With the ever increasing complexity of systems that our society deals with, the ab-initio understanding of important features remains a distant dream. However, one can distil many useful relations by simply collecting data about such complex systems and studying interdependencies. As our ability to gather, store and process data has rapidly progressed, the computational techniques to extract useful knowledge from data---machine learning (ML)---have become much more powerful. 
    One of the most popular ML techniques are neural networks (NNs), which have found numerous applications, from self-driving cars to drug discovery (see e.\,g. \cite{Wolf2018,Nielsen2015,Goodfellow2016}).
    
	Depending on the data, different learning scenarios of the ML algorithms can be distinguished. If the training data contains the desired outputs of the algorithm, the learning is called supervised.
	For example, for the task of image recognition the training data can be composed of images and corresponding labels.
	Sometimes, only partial knowledge about the desired output of the algorithm is available. For example, the full strategy of a game might be unknown while it is possible to assign a score to every set played. This is an example of semi-supervised or reinforced learning. Finally, often no labels are available with the data
	, and \textit{unsupervised} or \textit{self-supervised} \textit{learning} is applied.
	Autoencoders (AE) are a prominent example of NNs that learn without supervision, see e.\,g. \cite{Goodfellow2016}. They have been used, e.\,g., to denoise bird songs in the wilderness~\cite{BIRDpotamitis2016deep}.
     
    ML could benefit from the rapid progress of quantum computing quantum computing hard- and software (see e.\,g. \cite{Nielsen2010,Werner2001}).
    Moreover, there are important ML tasks where the data comes as a set of quantum, and possibly---highly entangled, states. Examples include quantum cryptography (see e.\,g. \cite{2019arXiv190601645P}), metrology (see e.\,g. \cite{RevModPhys.90.035005,giovannetti2011advances}), and chemistry (see e.\,g. \cite{szalay2015tensor,orus2019tensor}).
    ML is called quantum if it uses quantum algorithms or quantum data (see e.\,g. \cite{benedetti2019parameterized, Wikipedia2018}).

    Virtually every experimental preparation of a quantum state introduces noise.
	Usually, it is hard to design a denoising protocol. First, one has to identify and characterize all noise sources. Second, one has to invent a protocol which corrects the noise without affecting any relevant features of the quantum state. ML can automate this task. As there is often no denoised reference state to compare with, unsupervised learning is required.
   
    Various quantum neurons have been proposed in \cite{2019arXiv190210445B,Schuld2014, lewenstein_quantum_1994,wan_quantum_2017,DASILVA201655,2001arxiv7012A,NQ1008,KoudaMNP05,Torrontegui2018,FN18,SBS18,CGA17,KBA18,ABIMBK19,2018arXiv180810047S}. We follow \cite{2019arXiv190210445B}, since these NNs are capable of universal quantum computation, the computational complexity per training round scales only linearly with the depth of the NN, the cost function has a clear operational meaning, and the authors provide an open-source implementation. The parameters of such a quantum NN (QNN) are classical variables. In general, quantum parameters may be useful \cite{verdon2018universal}, but for ML tasks without memory they can give only a marginal improvement \cite{PhysRevLett.118.190503}.
            
    Classically simulable quantum AEs have been studied in~\cite{grant2018hierarchical}. In \cite{romero2017quantum} shallow quantum AEs have been introduced for data compression.
    The QNNs in \cite{romero2017quantum} are closely related to the neurons from \cite{2019arXiv190210445B}. Contrary to a claim in \cite{romero2017quantum} they are universal. However, the authors of \cite{romero2017quantum} restrict the class of operations to get polynomial complexity scaling with the width of the network. Data compression via AEs has been demonstrated with photons \cite{pepper2019experimental}. In \cite{lamata2018quantum}, it has been proposed to train AEs for quantum data compression using genetic algorithms on a classical computer. The trained AEs have been implemented on superconducting qubits \cite{ding2019experimental}. A quantum Boltzmann machine \cite{PhysRevX.8.021050} has been employed in a variational AE \cite{Khoshaman_2018} learning from classical data. Classical ML techniques have been used to design experiments that produce entangled states \cite{Melnikov1221,nichols2018designing} or useful entangled states robust against noise \cite{Knott_2016,nichols2018designing}. The general setting of quantum unsupervised ML has been studied in~\cite{sentis2019unsupervised}.
    
    In this work we 
    construct  quantum AEs capable of quantum advantage for the purpose of denoising quantum data. 
    We apply them to single and continuously parameterized sets of small highly entangled states subject to different kinds of noise. We observe excellent denoising without fine tuning of the hyperparameters.


\begin{figure}[t]
	\begin{center}
		\includegraphics[scale = 1]{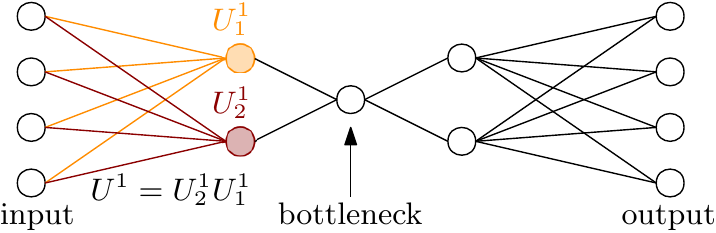}
	\end{center}
	\caption{Network architecture of an AE. The bottleneck prevents the AE from just copying the input data to the output so that it has to extract relevant features. Each neuron unitary acts on its qubit and the connected qubits in the previous layer (e.\,g. gold or red).}
	\label{fig:1}
\end{figure}

\section{Quantum autoencoders}
\label{sec:QAE}

\begin{figure*}[tb]
	\begin{center}
		\includegraphics[width=\columnwidth]{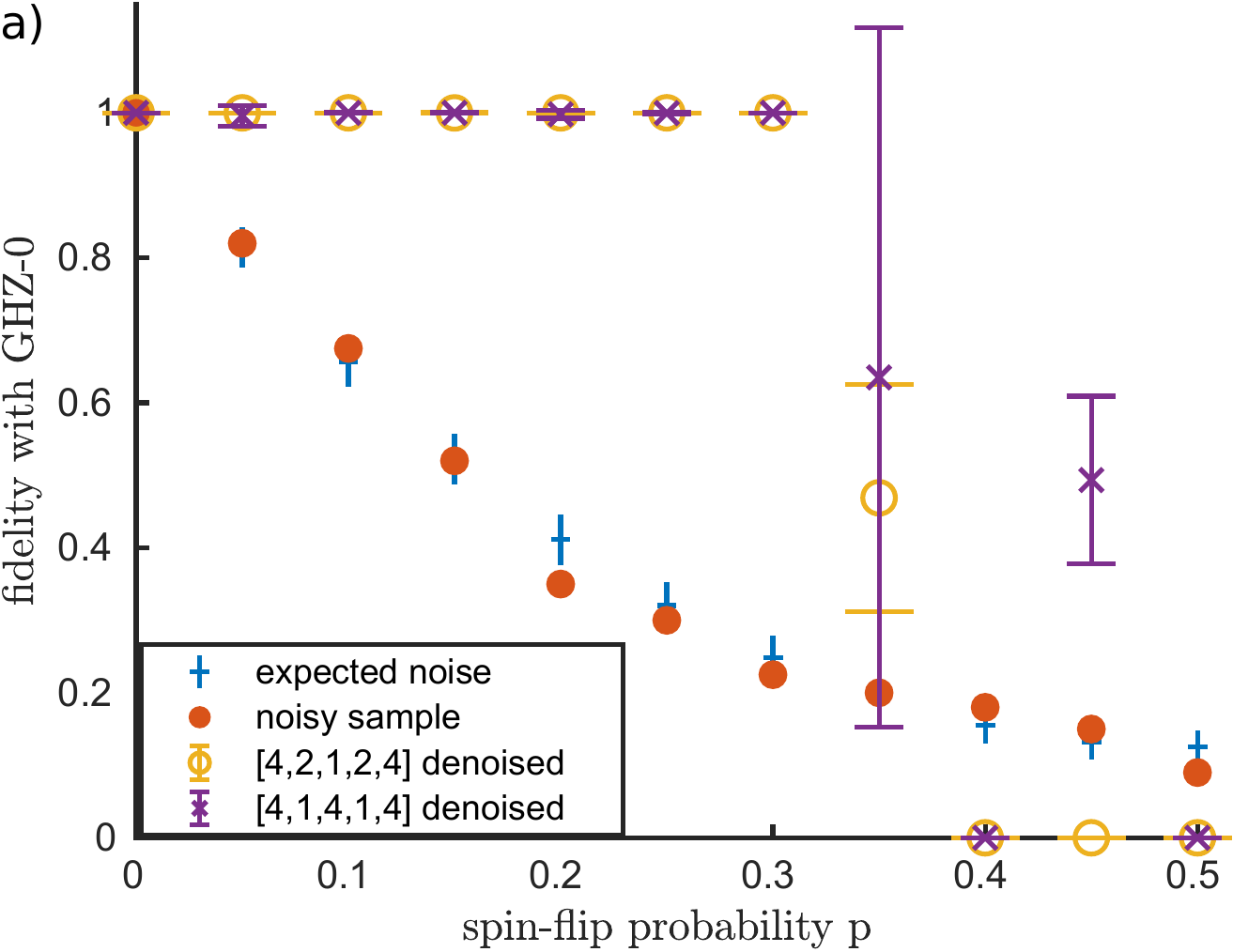}\includegraphics[width=\columnwidth]{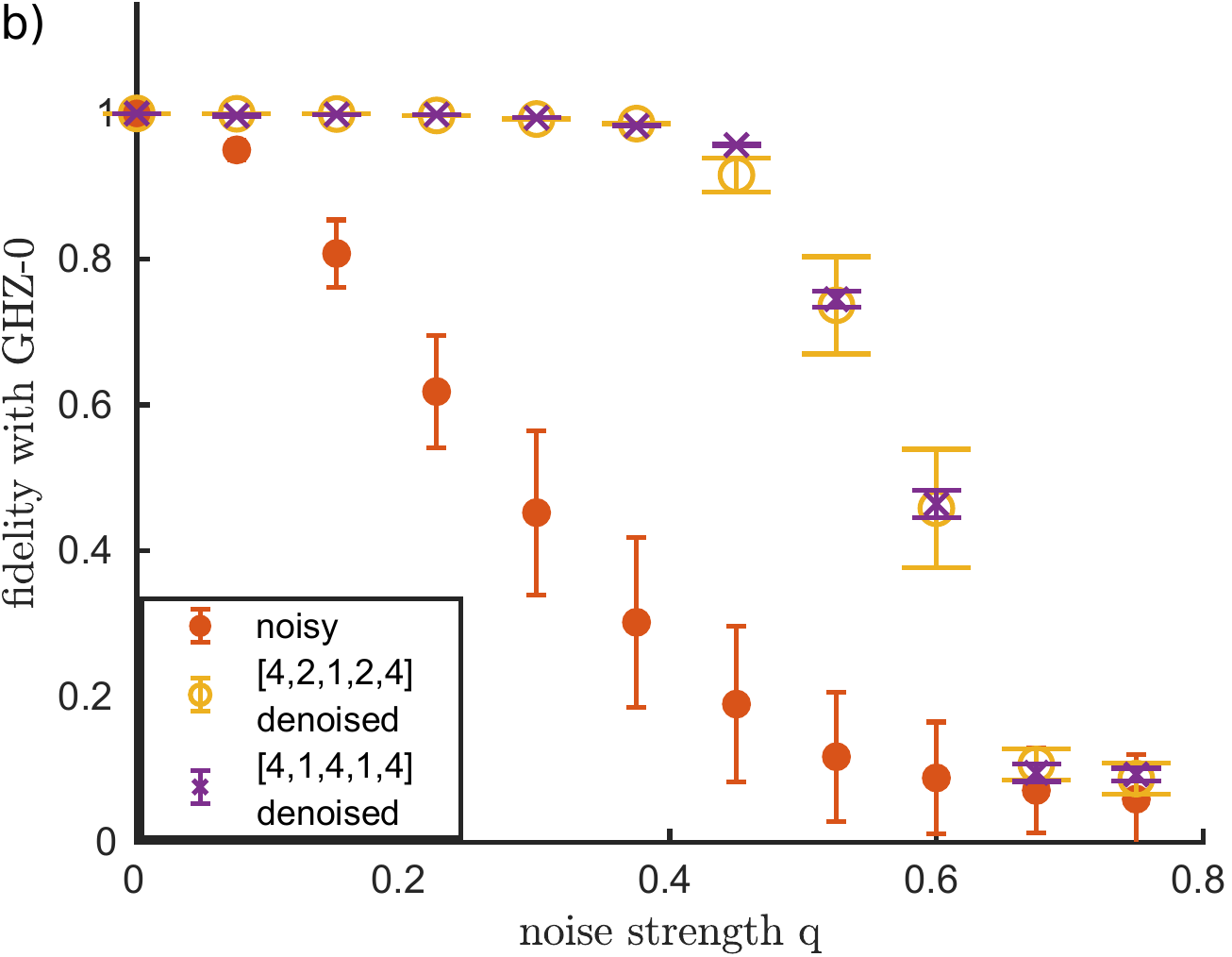}
	\end{center}
	\caption{Quantum AEs denoising GHZ-$0$ states. We show the average fidelity of noisy test states with the GHZ-$0$ state before denoising (red dots, $\bar F$) and after denoising (yellow circles / violet crosses, $\bar F_{\text{val}}$). Error bars display standard deviations. The arrays $[4,2,1,2,4]$ and $[4,1,4,1,4]$ indicate different AE topologies. 
	$200$ noisy training pairs, training rounds, and noisy test states per $p$ and $q$.
	(a) Correcting spin-flip errors. Blue plus signs show $\bar F^\infty\pm\Delta F^\infty$. (b) Correcting for random unitary noise.}
	\label{fig:2}
\end{figure*}
\begin{figure}
	\begin{center}
		\includegraphics[width=\columnwidth]{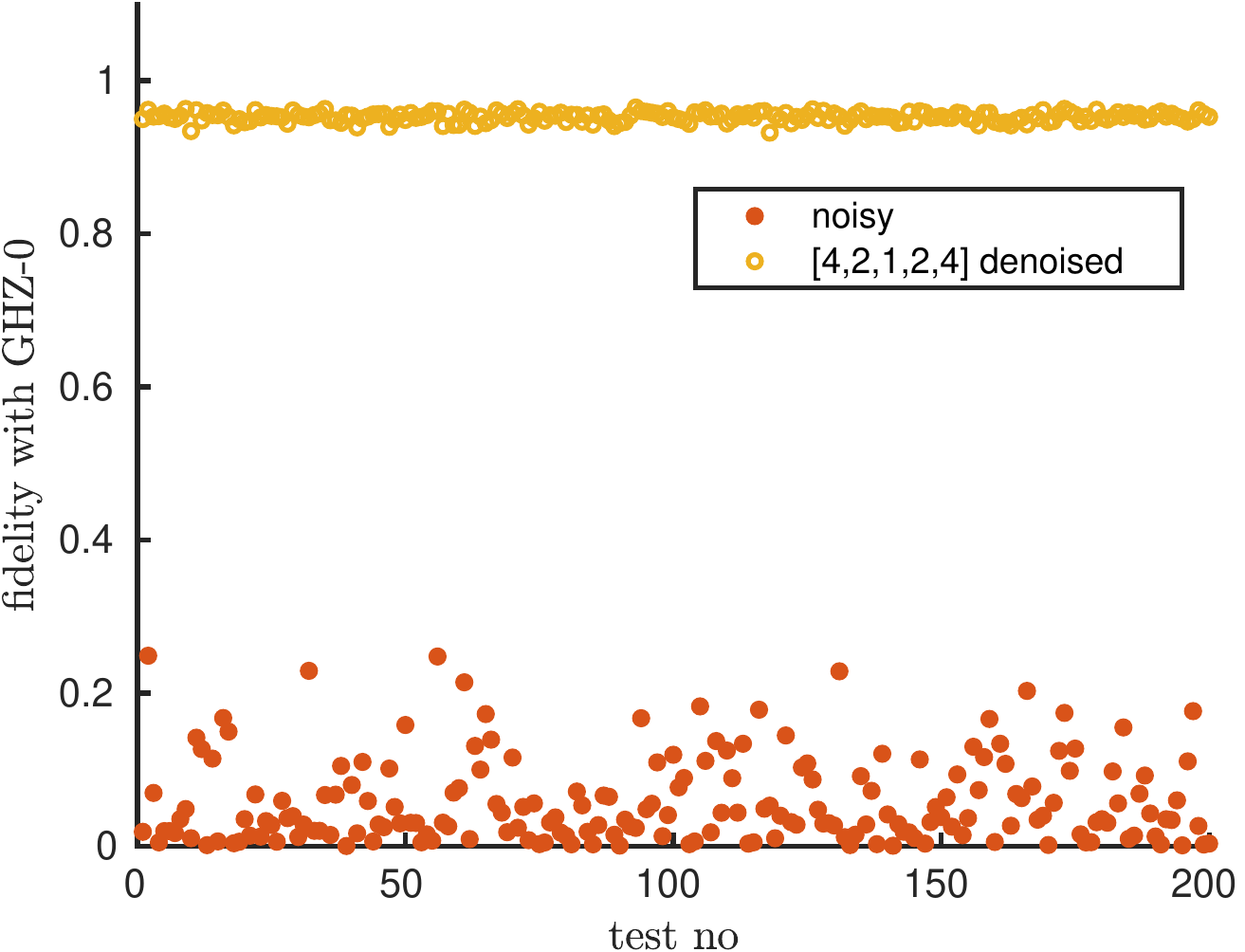}
	\end{center}
	\caption{A $[4,2,1,2,4]$ quantum AE correcting GHZ-$0$ states for combined noise: random unitary transformations with $q=0.3$ after spin-flip errors with $p=0.2$. We show the fidelity of each noisy test state with the GHZ-$0$ state before denoising (red dots, $F^{(i)}$) and after denoising (yellow circles, $F^{(i)}_{\text{val}}$). The respective average fidelities and standard deviations are $\bar F = \num{0,058}$, $\Delta F = \num{0,056}$, $\bar F_{\text{val}} = \num{0,953}$, and $\Delta F_{\text{val}} = \num{0,006}$.
	$200$ noisy training pairs and training rounds.} 
	\label{fig:3}
\end{figure}

	ML algorithms numerically solve variational problems. A NN is a variational class of maps $f_v:X\to Y$ parameterized by a vector $v$. It is constructed from simpler parameterized maps called neurons. The outputs of a set of neurons---a \textit{layer}---are fed into the next layer. If layer $n$ gets all its inputs from layers $k<n$, the network is called \textit{feed forward} (FF). The input $x\in X$ of the NN is the first layer, and the output $f_v(x)$---the last. The number of layers is the \textit{depth} of a NN, and the maximal number of neurons per layer---its \textit{width}. The geometry of the neuronal interconnections---the \textit{topology} of a NN---and the choice of neurons determine the variational class given by the NN. With suitable neurons any map can be represented as a FFNN (FFNN are \textit{universal}).
	
	Let us assume that a number of correct input-output pairs of the desired map---the \textit{training data} \mbox{$\{x_i,y_i \}_{i=1}^L \in X^L \times Y^L$}---is provided. The variational parameters in $v$ are optimized such that a \textit{cost function} \mbox{$C(\{x_i,y_i \}_{i=1}^L) = \frac{1}{L}\sum_{i=1}^L d(f_v(x_i),y_i)$} reaches a minimum. Here, $d$ is an appropriate distance measure. Typically the optimization employs some variant of the gradient descent algorithm (see e.\,g.~\cite{ruder2016overview}).
	 
	An AE is an FFNN  for extracting the most relevant features from the input data. The network has a bottleneck---a layer with smaller width than the (equal) input and output layers. The training data is a set $\{x_i,x_i \}_{i=1}^L$ of equal training inputs and reference outputs. In general, the desired output for $x$ is not $x$ itself: the bottleneck (see Fig.\ref{fig:1}) should force the AE to discard irrelevant information. Since no correct reference outputs are provided, the training of AEs is unsupervised.
	
	We specify the quantum neuron from \cite{2019arXiv190210445B} by attributing a single qubit to every neuron. Let $\{|\!\!\uparrow\rangle, |\!\!\downarrow\rangle\}$ denote an orthonormal basis of a qubit. In each layer following on the input, the $j$th neuron acts by a unitary $U_j$ on its own qubit and the preceding layer. The non-input qubits are initialized in $|\!\!\downarrow\rangle$. The $k$th layer, $k>1$, of $m$ neurons maps the state $\rho_{k-1}$ of layer $k-1$ onto
	\begin{equation}
	\label{LayerDef}
	    \mathcal N^k(\rho_{k-1}) \equiv \text{tr}_{k-1} 
	    \left(U 
	        \left(
	        \rho_{k-1} \otimes
	            (|\!\! \downarrow \rangle_{\text{out}} 
	            \langle \downarrow \!\!|)^{\otimes m}
	        \right)
	    U^{\dagger}\right),
	\end{equation}
	where the unitary $U \equiv U_m \dots  U_1 $ is subject to optimization (see Fig.~\ref{fig:1}). Note that this definition is related to the general form of a quantum channel (see e.\,g. \cite{2019arXiv190210445B, wolf2012quantum,2015arXiv150503106B,Werner2001}). 
	The quantum channel describing the full network with $M$ layers is
	\mbox{$
	    \mathcal{N}(\rho^{\text{in}}) = 
	    \mathcal{N}^M( \cdots \mathcal{N}^2
	    (\rho^{\text{in}}) \cdots ).
	$}
	Our distance measure is one minus the fidelity $F$. For training data $\{ \rho^{\text{in}}_i, | \psi^{\text{ref}}_i \rangle \}_{i=1}^{L}$ with pure desired outputs, $F(\rho,|\psi\rangle)=\langle\psi|\rho|\psi\rangle$ and the cost function reads
	\begin{equation}
 C \!\left(\{ \rho^{\text{in}}_i, | \psi^{\text{ref}}_i \rangle \}_{i=1}^{L} \right) = 
	    1-\bar F\!\left(\{\mathcal{N}(\rho^{\text{in}}_i ),|\psi^{\text{ref}}_i \rangle\}_{i=1}^{L}\right),
	\end{equation}
	where $\bar F(\{ \rho_i, | \psi_i \rangle \}_{i=1}^{L})=\frac{1}{L}\sum_{i=1}^{L}F(\rho_i,|\psi_i\rangle) \leq 1$.
	In the following, we abbreviate pure $\rho_i^{\text{in}}=|\psi_i^{\text{in}}\rangle\langle \psi_i^{\text{in}}|$ by $|\psi_i^{\text{in}}\rangle$.
	
	Due to the no-cloning theorem, it is impossible to use copies of the training inputs $|\psi_i^{\text{in}}\rangle$ as reference outputs $|\psi_i^{\text{ref}}\rangle$. Instead, these states have to be prepared independently. If the data source is noisy, the paired states will be different due to different noise realizations. However, if these states share essential features, the AE can still be trained.
	Below, we use half of the noisy training data as input and half as reference output in unsupervised learning.
	
	While, in practice, one has no access to the desired outputs of the NN $\{|\psi_i^{\text{id}} \rangle\}_{i=1}^{L}$, 
	the performance of AEs is best studied in a setting where these target states are known.
	We call the learning process successful if the mean \textit{validation function} 
	\begin{align}
	\bar F_{\text{val}}\!\left(\{ \rho^{\text{in}}_i, | \psi_i^{\text{id}} \rangle \}_{i=1}^{L}\right) =\bar F\!\left(\{ \mathcal{N}(\rho^{\text{in}}_i), | \psi_i^{\text{id}} \rangle \}_{i=1}^{L}\right)
	\end{align} is large, particularly, as compared to $\bar F(\{ \rho^{\text{in}}_i, | \psi_i^{\text{id}} \rangle \}_{i=1}^{L})$ before the NN is applied. 
	We define
	\begin{align}
	\begin{split}
	  F_{\text{val}}^{(i)}(\{\rho^{\text{in}}_i, | \psi_i^{\text{id}} \rangle \}_{i=1}^{L})&=F(\mathcal{N}(\rho^{\text{in}}_i),|\psi_i^{\text{id}}\rangle),\\ F^{(i)}(\{\rho^{\text{in}}_i, | \psi_i^{\text{id}} \rangle \}_{i=1}^{L}) &=F(\rho^{\text{in}}_i,|\psi_i^{\text{id}}\rangle).
	\end{split}
	\end{align}
	Note that the validation function, which compares  $\{  \mathcal{N}(\rho^{\text{in}}_i) \}_{i=1}^{L}$ with the target states, differs from the fidelity entering the cost function for training, which compares $\{  \mathcal{N}(\rho^{\text{in}}_i) \}_{i=1}^{L}$ with the noisy data.
	
    For the classical simulation of the quantum AE we have upgraded the MATLAB code from \cite{2019arXiv190210445B}. Most importantly, we now use the Nadam~\cite{Nadam,ruder2016overview} gradient descent algorithm. The updated code is available at~\cite{CODE}.


\section{Denoising GHZ states}
\label{sec:Noise}

\begin{figure*}[tb]
	\begin{center}
		\includegraphics[width=\columnwidth]{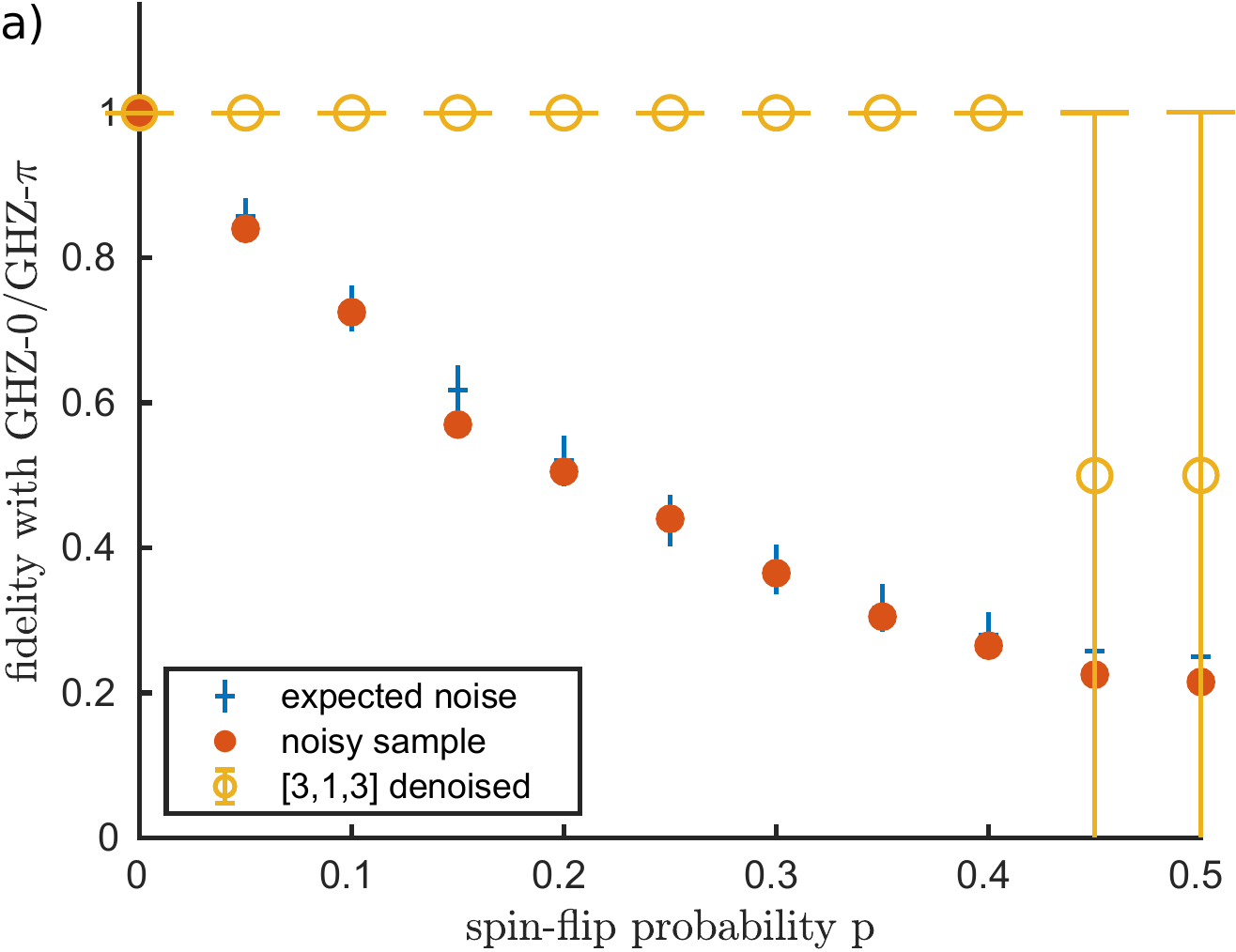}\includegraphics[width=\columnwidth]{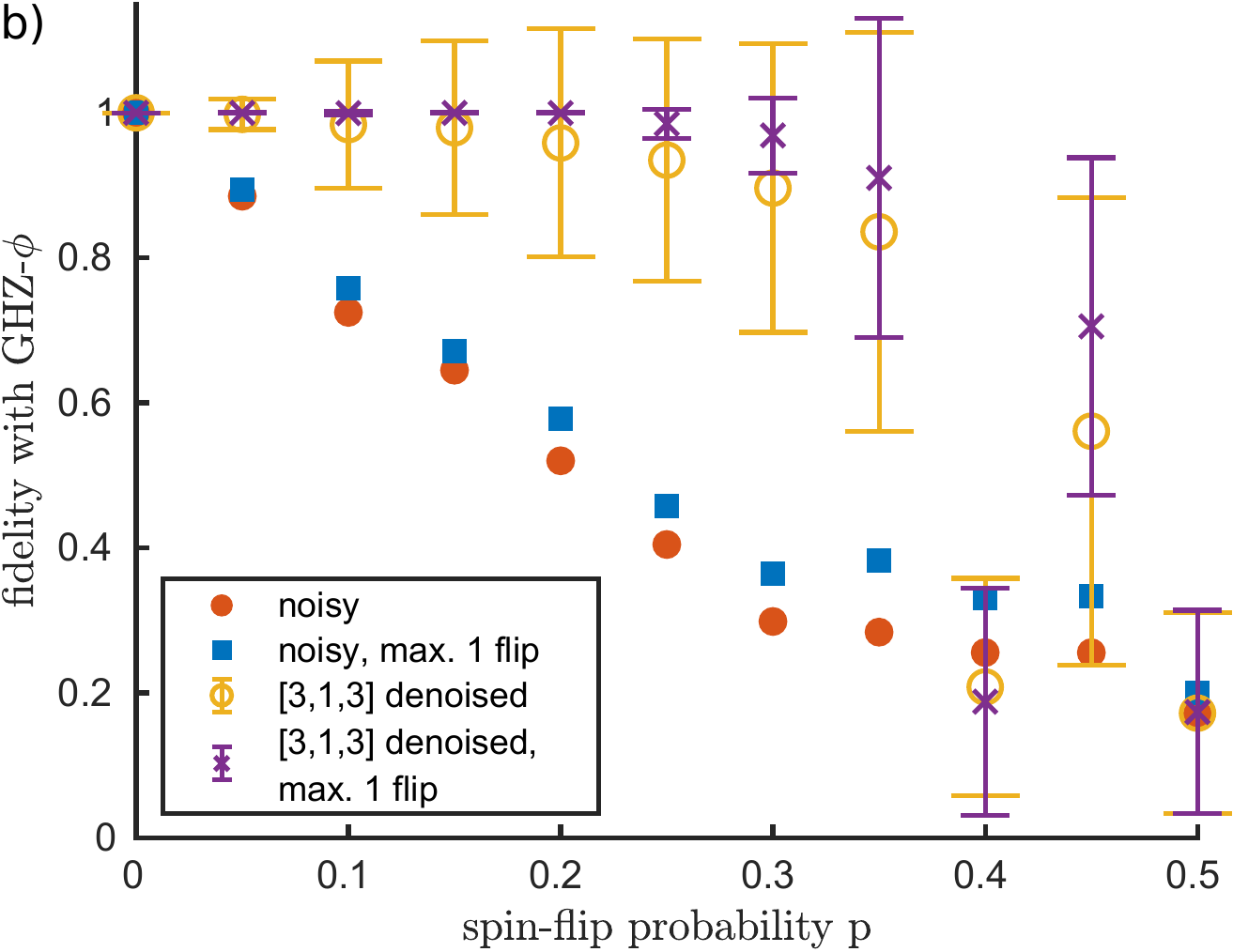}
	\end{center}
	\caption{$[3,1,3]$ quantum AEs correcting spin-flip errors in mixtures of GHZ-$\phi$ states with different phases $\phi$. We show the average fidelity of noisy test states with the respective noiseless GHZ-$\phi$ states before denoising (red dots) and after denoising (yellow circles). Error bars display standard deviations. For each $p$: $100$ training pairs per training phase, $200$ training rounds, and $200$ test pairs. (a) Fifty-fifty mixture of GHZ-$0$ and GHZ-$\pi$ states, both for training and testing. Blue plus signs show $\bar F^\infty\pm\Delta F^\infty$. (b) Training phases $\{0,\pi/3,2\pi/3,\pi\}$, and testing on random phases $\phi\in(0,\pi)$. Blue squares before denoising and violet crosses after denoising are obtained for the test states with $|J|\leq 1$.}
	\label{fig:4}
\end{figure*} 

We call
\begin{equation}
|\!\operatorname{GHZ}_\phi\rangle = \frac{1}{\sqrt{2}}\left(|\!\!\uparrow\rangle^{\otimes m}+\operatorname{e}^{i\phi}|\!\!\downarrow\rangle^{\otimes m}\right)
\end{equation}
an $m$-qubit GHZ state with phase $\phi$ or a GHZ-$\phi$ state. GHZ states have a maximal entanglement depth. This makes them an appealing resource for quantum information and quantum enhanced metrology. However, to fully exploit this resource the GHZ state has to be protected from experimental noise. In this section we show how quantum AEs can be used to denoise small GHZ states. We investigate two complementary noise processes: spin-flip errors and small random unitary transformations (see e.\,g. \cite{2015arXiv150503106B, Nielsen2010, wolf2012quantum}).

For spin-flip errors we assume that for a time $T$ all qubits are flipped back and forth at some rate $\Gamma$.  Thus each qubit has a probability of $p=(1-\operatorname{e}^{-2\Gamma T})/2\leq 0.5$ to end up in a flipped state. The flips of the $j$th qubit affect the density matrix $\rho$ of the initial, noiseless, $m$-qubit state according to
\begin{equation}
\mathcal{E}_j(p,\rho)= p \sigma^x_j\rho \sigma^x_j+(1-p)\rho,\;\sigma^x_j=\bigotimes_{1}^{j-1}\! {I\!d} \otimes \sigma^x \bigotimes_{j+1}^m\! {I\!d}
\end{equation}
where ${I\!d}=|\!\!\uparrow\rangle\langle\uparrow\!\!|+|\!\!\downarrow\rangle\langle\downarrow\!\!|$ is the identity and $\sigma^x=\mbox{$|\!\!\uparrow\rangle\langle\downarrow\!\!|$}+\mbox{$|\!\!\downarrow\rangle\langle\uparrow\!\!|$}$ the spin-flip operator for a single qubit. The total noise channel is obtained by concatenating $\mathcal{E}_j$ for all qubits $j\in\{1,\ldots,m\}$:
\begin{equation} \label{tatatnoise}
\mathcal{E}(p,\rho)=\mathcal{E}_m(p,\mathcal{E}_{m-1}(p, \cdots \mathcal{E}_1(p,\rho)\cdots))
\end{equation}
We assume that in each experimental shot a subset $J\subseteq\{1,2,\ldots,m\}$ of a total of $m$ qubits is flipped. The probability of  $\rho_J=\prod_{j\in J}\sigma^x_j\rho \prod_{j\in J}\sigma^x_j$ is $P_p(J)=p^{|J|}(1-p)^{m-|J|}$. Note that states $\rho_J$ with different $J$ may coincide or have non-orthogonal supports.

For unitary noise we assume that a state evolves with a random time-dependent Hamiltonian. The noise strength is captured by a dimensionless parameter $q$. See Appendix~\ref{app:nm} for details.

{\it Denoising a GHZ state with zero phase.---} First, we show how well an AE can denoise $4$-qubit GHZ states with zero phase. We employ two AE topologies. One is the deep QNN
\raisebox{-.25\height}{\includegraphics[height=13pt]{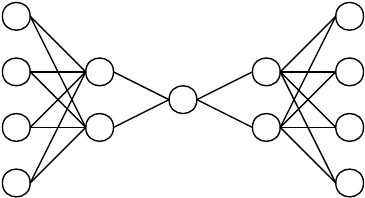}} denoted by $[4,2,1,2,4]$ and the other one is a stacked QNN: we train the AE
\raisebox{-.25\height}{\includegraphics[height=13pt]{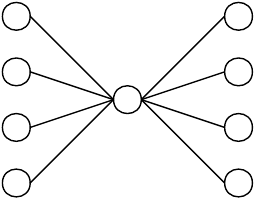}} $\sim [4,1,4]$ 
but denoise with
\raisebox{-.25\height}{\includegraphics[height=13pt]{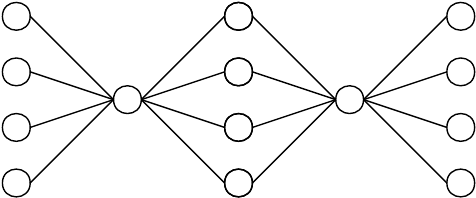}} $\sim [4,1,4,1,4]$ 
by applying $[4,1,4]$ twice.
Each training employs $200$ training pairs and takes $200$ steps of the gradient descent algorithm ($200$ {\it training rounds}). We test the trained AEs on $200$ GHZ-$0$ states exposed to the respective noise. The validation function, which, ideally, should reach one, is the fidelity between the denoised output of the AE and the GHZ-$0$ state.

Fig.~\ref{fig:1}a summarizes our results in the case of spin-flip errors. For each spin-flip probability $p$ we, first, draw the training data and one set 
of $L=200$ noisy test states $\{|\psi_i\rangle\}_{i=1}^L$ according to the probability distribution $P_p(J)$. We independently train both AE topologies. 
For each topology, we apply the respective AE to every $|\psi_i\rangle$ and get outputs $\rho_i$. To assess the performance of the AE, we evaluate the mean validation function after denoising---$\bar F_{\text{val}}(\{|\psi_i\rangle,|\!\operatorname{GHZ}_0\rangle \}_{i=1}^L)$
(yellow circles / violet crosses)---and compare it to its value before denoising---$\bar F(\{|\psi_i\rangle,|\!\operatorname{GHZ}_0\rangle \}_{i=1}^L)$
(red dots). We find that up to $p=0.3$ both AE topologies remove the spin-flip errors almost ideally, see Appendix~\ref{app:lod} for a discussion. 

The error bars of $\bar F_{\text{val}}$ indicate the standard deviation $\Delta F_{\text{val}}=\sqrt{\sum_i(F_{\text{val}}^{(i)}-\bar F^{\hphantom{(i)}}_{\text{val}})^2}$. Note that, contrary to $F_{\text{val}}^{(i)}$, $\bar F_{\text{val}}+\Delta F_{\text{val}}$ can exceed one. For the input, $\Delta F=\sqrt{\bar F(1-\bar F)}$ is large since $F^{(i)}\in\{0,1\}$. Instead of adding error bars to $\bar F$, we show how $\{|\psi_i\rangle\}_{i=1}^L$ compares to the ideal probability distribution $P_p(J)$ of spin-flipped GHZ-$0$ states.  The blue plus signs mark the expectation value of $F$, $\bar F^{\infty}= (1-p)^4+p^4$. Their vertical bars indicate the standard deviation $\Delta F^{\infty}/\sqrt{L}=\sqrt{\bar F^{\infty}(1-\bar F^{\infty})/L}$, which characterizes the spread of the average $\bar F(\{|\psi_i\rangle,|\!\operatorname{GHZ}_0\rangle \}_{i=1}^L)$ for independent draws of $L$ noisy states. 

Random unitary noise gives Fig.~\ref{fig:1}b. To get a train or test state, we evolve the GHZ state with a random unitary drawn according to the respective noise strength~$q$. We, again, compare the outcomes of the validation function before and after denoising. This time, we add error bars of size $\Delta F_{\text{(val)}}=\sqrt{\sum_i(F_{\text{(val)}}^{(i)}-\bar F^{\hphantom{(i)}}_{\text{(val)}})^2}$ to both $\bar F$ and $\bar F_{\text{val}}$. Virtually perfect denoising succeeds up to a noise strength of $q=0.375$.

Finally, we combine the two noise models. Spin-flip errors with $p=0.2$ are followed by random unitary transformations with $q=0.3$. We train and test the $[4,2,1,2,4]$ AE on the combined noise. Fig.~\ref{fig:3} shows that the AE impressively increases the fidelity of each noisy test state with the GHZ state.

{\it Denoising GHZ states with variable phase.---}
So far we have demonstrated that an AE can denoise the state on which it has been trained. But it can do better. An AE can learn to denoise multiple target states, including ones not contained in the training data. It is crucial, though, that the noise process is sufficiently different from the transformations connecting the target states. Otherwise, the attribution of noisy states to target states becomes ambiguous. Assume that an experiment encodes some information into the phase of a GHZ state, and that this GHZ state is affected by spin-flip errors. We show that an AE can denoise the output of such an experiment. We consider $3$-qubit states and employ the simplest possible AE topology:  \raisebox{-.25\height}{\includegraphics[height=13pt]{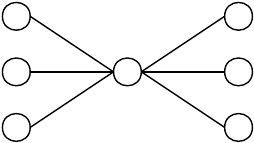}} $\sim [3,1,3]$.

As a first example, we imagine an experiment which outputs either a GHZ state with zero phase, GHZ-$0$, or with phase $\pi$, GHZ-$\pi$. The AE is trained for $200$ rounds on $100$ pairs of noisy GHZ-0 states and $100$ pairs of noisy GHZ-$\pi$ states. To test the performance of the trained AE, we apply it to $100$ noisy GHZ-0 states and $100$ noisy GHZ-$\pi$ states and compare each output to the respective noiseless state. Fig.~\ref{fig:4}a shows that the AE excellently denoises the two GHZ states up to a spin-flip probability of $p=0.4$. Note that the AE deduces whether the experiment has given a phase of zero or $\pi$ from the particular noisy input state alone.

Our second example is even more demanding. We assume that the experiment can output a GHZ state with any phase $\phi\in[0,\pi]$. We restrict the phase to $[0,\pi]$ because it is impossible to distinguish a GHZ-$\phi$ state with $|J|$ flipped qubits from a GHZ-$(-\phi)$ state with $3-|J|$ flipped qubits. The training involves only four equidistant training phases $\phi_i$ between $\phi_0=0$ and $\phi_4=\pi$. It, again, employs $100$ training pairs per $\phi_i$ and takes $200$ training rounds. We test the AE on $200$ noisy GHZ-$\phi$ states with randomly chosen phases $\phi\in(0,\pi)$.

Considering a GHZ-$\phi$ state with $\phi\notin\pi\mathbb{Z}$ roughly doubles the number of different spin-flipped states as compared to $\phi\in\pi\mathbb{Z}$. 
Only for $\phi\in\pi\mathbb{Z}$, the flipped $m$-qubit states $\prod_{j\in J}\sigma^x_j|\!\operatorname{GHZ}_\phi\rangle$ and $ \prod_{j\in M\setminus J}\sigma^x_j|\!\operatorname{GHZ}_\phi\rangle$ with $J\subseteq M=\{0,1,\ldots,m\}$ are, up to a global phase, identical. As a consequence, for $3$-qubit GHZ-$\phi$ states with $\phi\in\pi\mathbb{Z}$, correcting spin-flip errors with $|J|=1$ suffices for perfect denoising. For $\phi\notin\pi\mathbb{Z}$, errors with $|J|=2$ and $|J|=3$ need to be regarded separately.

Fig.~\ref{fig:4}b displays the capability of the AE to denoise GHZ states with a random phase. Note that for $p=0$ the fidelity of the outputs with the test states reaches one. Because of the bottleneck, the AE cannot learn the identity operation; nevertheless it correctly reproduces GHZ states with phases not contained in the training data. The AE improves the average value of the validation function for $p\leq 0.35$, but it leaves a considerable variance (yellow circles). However, if we keep only the test states with $|J|\leq 1$, we observe excellent denoising up to $p=0.2$ (violet crosses).

\FloatBarrier
\section{Discussion}
\label{sec:End}

We have constructed quantum AEs 
and have shown that these AEs can remove spin-flip errors and random unitary noise from small GHZ states. Particularly, correcting spin-flip errors has succeeded for a set of GHZ states parameterized by a continuous phase parameter. Thus, AEs for denoising can be used not only for state preparation but also for metrology.
In principle, our method can be applied to any set of quantum states subject to any kind of noise. 
Further possible applications of quantum AEs include data compression, quantum error correction, and parameterized state preparation. 

We expect that larger input states will require deeper networks. The number of quantum gates needed for one application of the fully connected AE scales exponentially with the width but only linearly with the depth of the network. The exponential scaling can be avoided by constraining the QNN, e.\,g., using sparse networks as in Appendix~\ref{app:sparse}.

Small universal quantum computers have been realized on several physical platforms, e.\,g. superconducting qubits and trapped ions \cite{qcSupercond,qcIons}.
If the state to be denoised is prepared on the same platform as the AE, both may be affected by equal noise, and the AE may become too noisy for denoising. However, there is a great interest in hybrid systems, which have been demonstrated, e.\,g., for superconducting qubits coupled to atomic and spin ensembles and for trapped ions with cold atoms \cite{hybridSuperconducting2,hybridSuperconducting,hybridIon}. Our proposal can help to denoise states from a noisy platform using a well-controlled one, or to remove deteriorating effects introduced at the interface between the coupled platforms. The impact of noise affecting the AE itself will be discussed in~\cite{BachelorD}.

Training an AE requires much more computational resources than testing it. To approach the experimental implementation, a small AE trained on a classical computer can be tested on a quantum computer, as has been done in \cite{ding2019experimental} for data compression. Moreover, the photonic realization~\cite{pepper2019experimental} of a compressing quantum AE suggests that also the training of our AE is within the reach of current quantum technology.


\begin{acknowledgments}
We thank Abhinav Anand, Kerstin Beer, Terry Farrelly, Carsten Klempt, Taras Kucherenko, Tobias J. Osborne, Robert Salzmann, Luis Santos, Augusto Smerzi, Dorothee Tell, and Ramona Wolf for useful discussions. We acknowledge support by the SFB 1227 ``DQ-mat'', projects A02 and A06, of the German Research Foundation (DFG). 
\end{acknowledgments}

\FloatBarrier
\appendix

\section{Unitary noise}
\label{app:nm}

For unitary errors we assume that any one-qubit error can occur with probability $p_u$ on every qubit. The noise channel for the $j$th qubit can be written as
\begin{align}
\mathcal{E}_j(p_u,\rho)=&\, p_u \text{tr}_{m+1} \! \left[ \text{SWAP}_{j,m+1} \left(\rho \otimes \frac{{I\!d}}{2} \right) \text{SWAP}^{\dagger}_{j,m+1} \right] \nonumber \\& + (1-p_u)\rho
\end{align}
and the total noise channel is obtained by concatenating $\mathcal{E}_j$ for all qubits (\ref{tatatnoise}).
Here $\text{tr}_j(\cdot)$ is a partial trace and $\text{SWAP}_{j,m}$ swaps the $j$th with the $m$th qubit. 
A random one-qubit error can be attributed to the evolution with a random time-dependent Hamiltonian $H(t)$, where the error probability $p_u$ is a monotonic function of the interaction strength and the evolution time $T$. The evolution with $H(t)$ can be constructed using a quantum Brownian circuit \cite{zhou2019operator,Lashkari2013}. We consider a family of Hamiltonians $\{H_j = H(j\Delta t)\}_{j=1}^{n}$,
$\Delta t = T/n$ 
such that the entries of every Hermitian $H_j$ are Gaussian distributed with zero mean and a standard deviation of $\frac{2\pi \hbar \nu}{\sqrt{2^m n}}$. The dimensionless parameter $q=\nu T$ captures the noise strength.
We assume that in each experimental shot the initial state evolves
with the unitary operator
\begin{equation}
U = \prod_{j=1}^{n} \exp\!\left( i H_j\Delta t/\hbar \right).
\label{eq:randU}
\end{equation} 
By It\^{o}'s calculus, there exists $H(t)$ such that
\begin{equation}
U 
=
\mathcal{T}\exp\!\left( i/\hbar \int_{0}^{T}\! H(t) \operatorname{d}\!t\right) + O\left(\frac{1}{\sqrt{n}}\right),
\end{equation} 
where $\mathcal{T}$ is the time ordering operator.
We use $n=20$.

\begin{figure}[b]
	\begin{center}
		\includegraphics[width=\columnwidth]{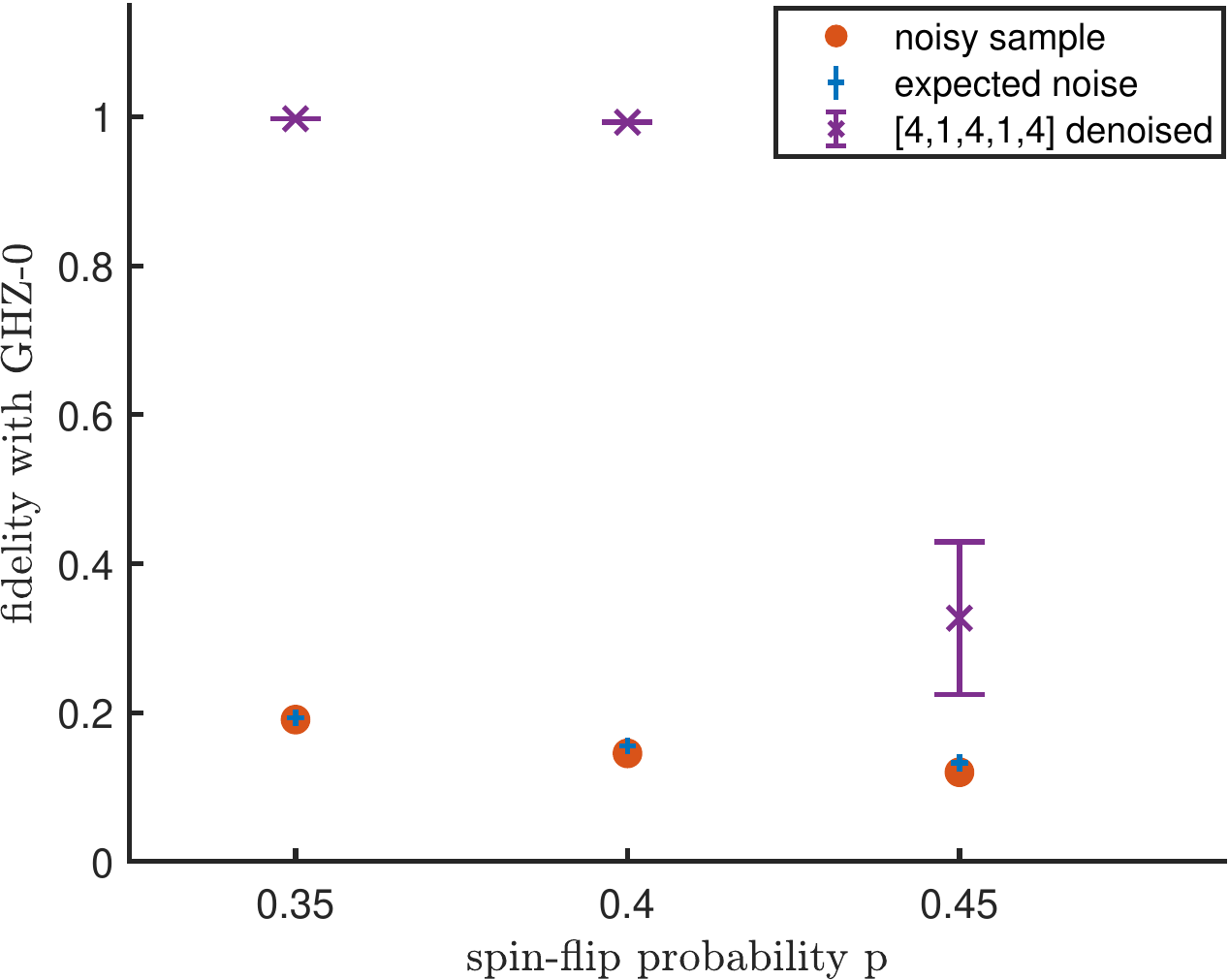}
	\end{center}
	\caption{A stacked $[4,1,4,1,4]$ quantum AE removing spin-flip errors from the GHZ-$0$ state.
	 We show the average fidelity of noisy test states with the GHZ-$0$ state before denoising (red dots, $\bar F$) and after denoising (violet crosses, $\bar F_{\text{val}}$). Error bars display standard deviations. Blue plus signs show $\bar F^\infty\pm\Delta F^\infty$.
	$1500$ noisy training pairs, $1500$ noisy test states, and $75$ training rounds per $p$.  
	}
	\label{fig:5}
\end{figure}
\section{Limitations of denoising}
\label{app:lod}

A simple argument suggests that an AE might denoise GHZ-$0$ states all the way up to $p=0.5$.
For all $p<0.5$ the most probable of all spin-flipped GHZ-$0$ states is the GHZ-$0$ state itself. All (non-identical) flipped GHZ-$0$ states are orthogonal to each other. Hence, the state $\rho$ which maximizes the average fidelity with the ideally distributed flipped GHZ-0 states, $\operatorname{argmin}_\rho \sum_JP_p(J)\langle \operatorname{GHZ}_0\!|\prod_{j\in J}\mathcal{F}_j\rho\prod_{j\in J}\mathcal{F}_j|\!\operatorname{GHZ}_0\rangle$, is the original, noiseless GHZ-$0$ state. 

Why do our AEs fail to denoise GHZ-$0$ states beyond $p\approx 0.3$? For $p\rightarrow 0.5$ all flipped GHZ states become equally probable. On a finite training sample the ordering by probability can get misrepresented. Furthermore, the small difference in cost corresponding to a small difference in probabilities can be missed due to a finite training precision. Finally, an actual AE can learn more than a single state. Note also that, so far, we have not optimized the hyperparameters of the gradient descent for individual noise strengths.

For Fig.~\ref{fig:5} we, again, train our stacked $[4,1,4,1,4]$ AE to remove spin-flip errors from the GHZ-$0$ state. As compared with Fig.~\ref{fig:1}a, we increase the number of training pairs from $200$ to $1500$ and optimize the gradient descent for a large spin-flip probability $p$. As a result, denoising succeeds up to $p=0.4$ instead of $p=0.3$.
\section{Sparse networks}
\label{app:sparse}
\begin{figure}[t]
	\begin{center}
		\includegraphics[scale=1]{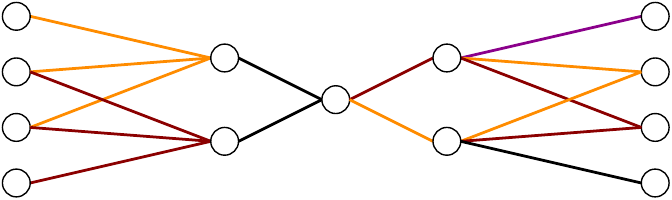}
	\end{center}
	\caption{A sparse $[4,2,1,2,4]$ quantum AE. In each layer, different colors highlight different neurons.
	}
	\label{fig:6}
\end{figure}
%
\begin{figure*}[bt]
	\begin{center}
		\includegraphics[width=\columnwidth]{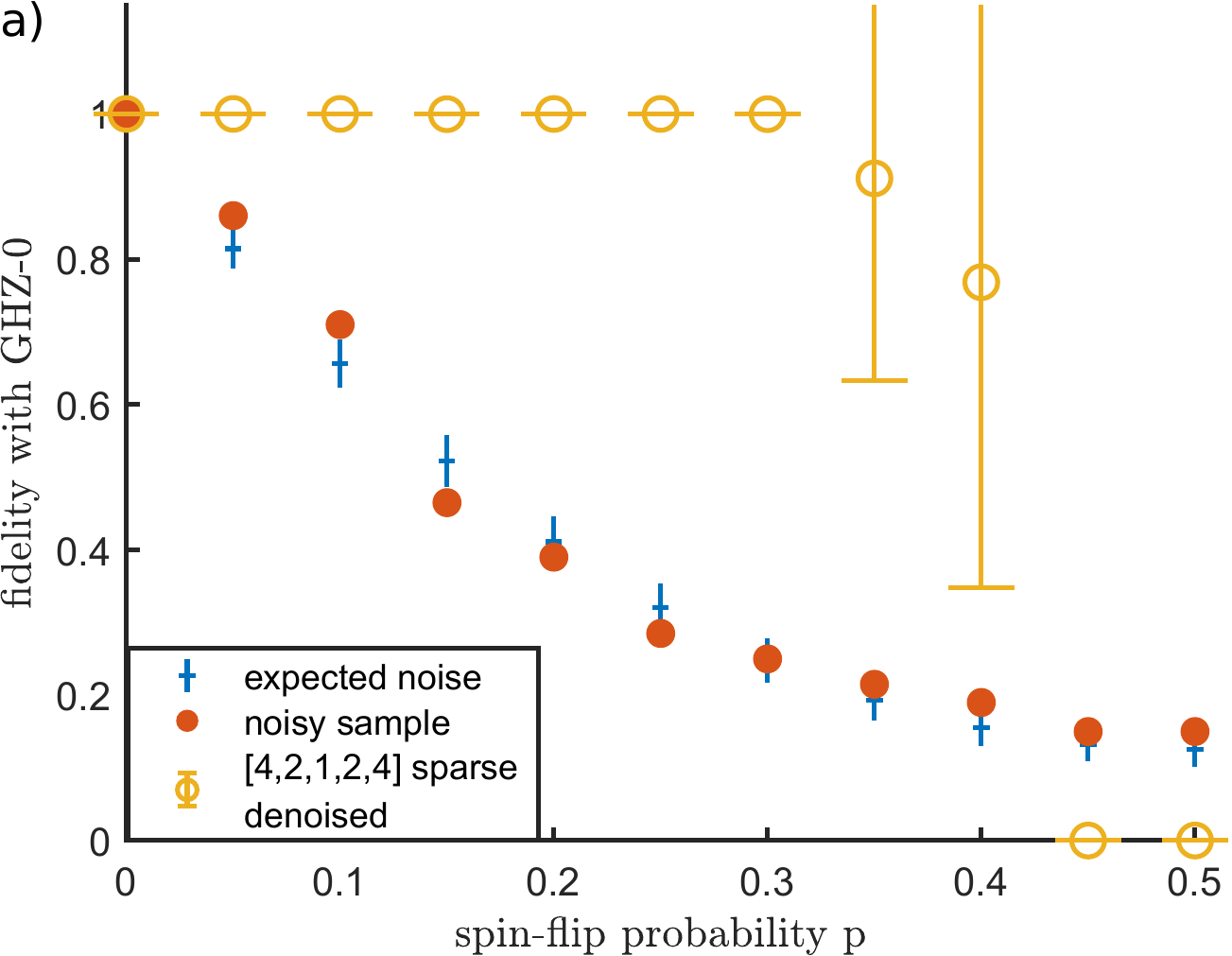}\includegraphics[width=\columnwidth]{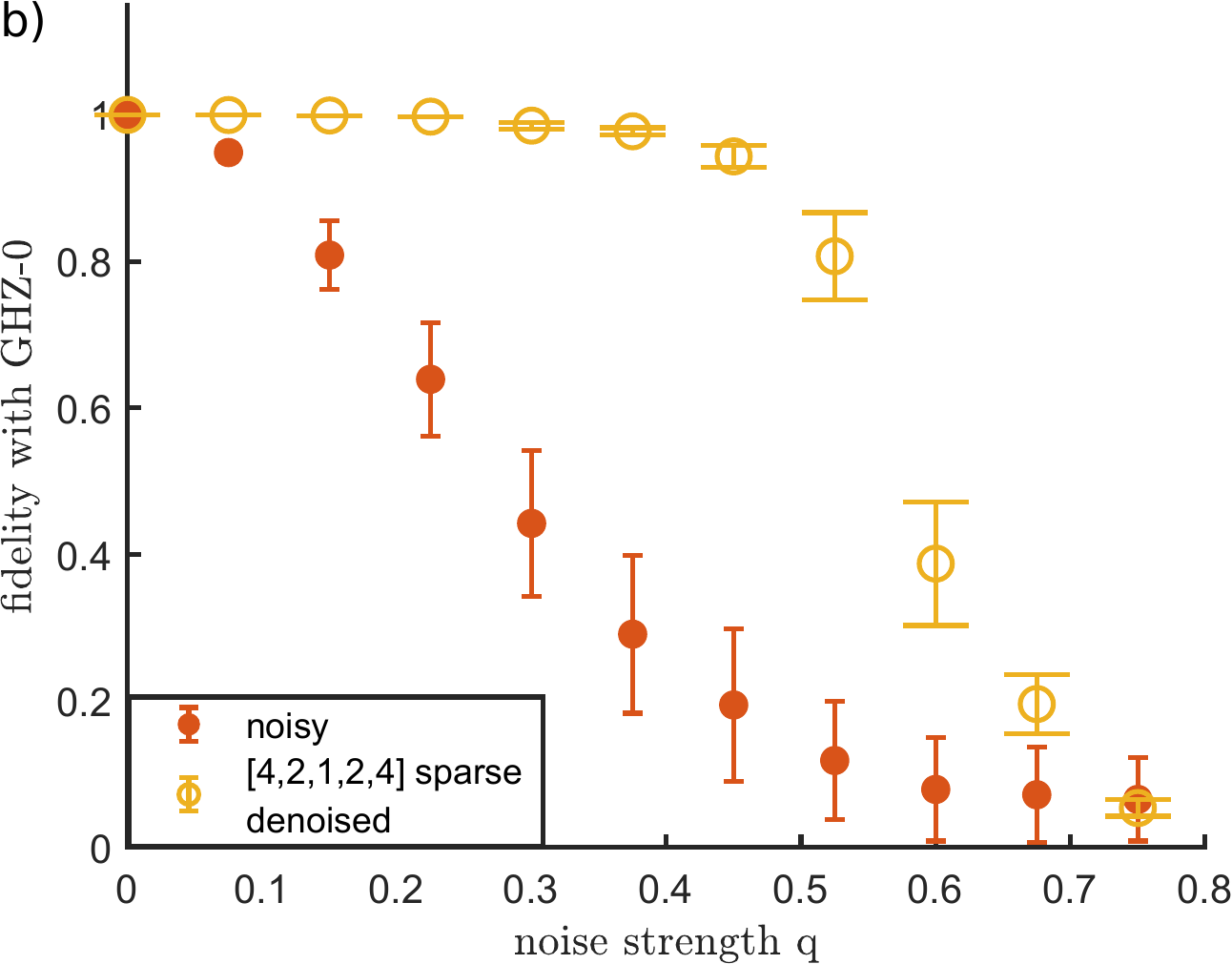}
	\end{center}
	\caption{Sparse $[4,2,1,2,4]$ 
	quantum AE (see Fig.~\ref{fig:6}) denoising GHZ-$0$ states. We show the average fidelity of noisy test states with the GHZ-$0$ state before denoising (red dots, $\bar F$) and after denoising (yellow circles, $\bar F_{\text{val}}$). Error bars display standard deviations. $200$ noisy training pairs, training rounds, and noisy test states per $p$ and $q$.
	(a) Correcting spin-flip errors. Blue plus signs show $\bar F^\infty\pm\Delta F^\infty$. (b) Correcting for random unitary noise.}
	\label{fig:7}
\end{figure*}

In general, a neuron does not have to be connected to all the neurons in the preceding layer. A NN containing neurons with less connections is called sparse. Sparse networks depend on less variational parameters. On one hand, this reduces the variational class. Eventually, such a QNN may become classically simulable (see e.\,g.~\cite{grant2018hierarchical}). On the other hand, this speeds up both training and application, and makes the network less prone to overfitting. Recall that the number of gates needed for one application of a fully connected network scales exponentially with its width. If the number of connections per neuron is kept constant, this scaling becomes linear.

We observe that full connectivity is not essential for the success of our quantum AE. Fig.~\ref{fig:7} shows the denoising capability of the sparse $[4,2,1,2,4]$ AE depicted in Fig.~\ref{fig:6}. The results turn out to be compatible with the corresponding fully connected network, see Fig.~\ref{fig:2}. To the advantage of experiments, our sparse topology is local---the retained connections are immediately adjacent.

%
\FloatBarrier
\bibliography{autoencoder.bib}

\end{document}